\documentclass[final, 3p, sort&compress]{elsarticle} 

\usepackage[T1]{fontenc}
\usepackage[utf8]{inputenc}
\usepackage[english]{babel}
\usepackage{amsmath}
\usepackage{amssymb}
\usepackage{lipsum}
\usepackage{adjustbox}
\usepackage{color}
\usepackage[percent]{overpic}
\usepackage{float}
\usepackage[caption = false]{subfig}
\usepackage{lineno}
\usepackage{orcidlink} 
\usepackage{etoolbox}
\usepackage{multicol} 
\usepackage{dblfloatfix}

\allowdisplaybreaks

\makeatletter
\def\ps@pprintTitle{%
  \let\@oddhead\@empty
  \let\@evenhead\@empty
  \def\@oddfoot{\reset@font\hfil\thepage\hfil}
  \let\@evenfoot\@oddfoot
}
\makeatother

\begin{document}

\begin{frontmatter}



\title{First measurement of polarized spin-density matrix elements and differential cross sections d$\sigma$/d$t$
  in $\omega$~photoproduction off the proton for $2.7 < E_\gamma <  5.2$~GeV using CLAS at Jefferson Lab}
  
\author{

\begin{small}
\begin{center}
T.~Hu$^{\,14}$,
Z.~Akbar$^{\,14}$,
V.~Crede$^{\,14,\,\ast}$,
J.M.~Laget$^{\,43}$,
V.~Mathieu$^{\,54,\,55}$,
M.J.~Amaryan$^{\,35}$,
W.R.~Armstrong$^{\,1}$,
H.~Atac$^{\,42}$,
N.A.~Baltzell$^{\,41,\,43}$,
L.~Barion$^{\,17}$
% ...
\\
\vspace{0.2cm}
The CLAS Collaboration at Jefferson Laboratory\\
\vspace{0.2cm} {\it
  $^{1}$\,Argonne National Laboratory, Argonne, Illinois 60439, USA\\
% ...
$^{14}$\,Florida State University, Tallahassee, Florida 32306, USA\\
% ...
$^{17}$\,INFN, Sezione di Ferrara, 44100 Ferrara, Italy\\
% ...
$^{35}$\,Old Dominion University, Norfolk, Virginia 23529, USA\\
% ...
$^{41}$\,University of South Carolina, Columbia, South Carolina 29208, USA\\
  $^{42}$\,Temple University, Philadelphia, PA 19122, USA\\
  $^{43}$\,Thomas Jefferson National Accelerator Facility, Newport News, Virginia 23606, USA\\
% ...
$^{54}$\,Departamento de F\'isica Te\'orica, Universidad Complutense de Madrid and IPARCOS,\\ E-28040 Madrid, Spain\\
  $^{55}$\,Departament de F\'isica Qu\`antica i Astrof\'isica and Institut de Ci\`encies del Cosmos,\\
                 Universitat de Barcelona, E-08028, Spain\\[1ex]
}\end{center}
$\ast$~Corresponding author: crede@fsu.edu            
\vspace{0.1cm}
\end{small}
} 

\begin{abstract}
We report on the differential cross sections d$\sigma$/d$t$, the unpolarized spin-density matrix elements
$\rho^0_{00}$, $\rho^0_{1-1}$, Re\,$\rho^0_{10}$, and the first extraction of the polarized elements
Im\,$\rho^3_{10}$, Im\,$\rho^3_{1-1}$ for the reaction $\gamma p\to p\omega$ using the CLAS spectrometer at
Jefferson Laboratory. The $\omega$~mesons were detected in their dominant charged decay mode, $\omega \to
\pi^+\pi^-\pi^0$, and all $t$-dependent results are presented in a fine binning for incident photon energies
between 2.73 and 5.16~GeV (corresponding to the center-of-mass energy range $W \in [\,2.45,3.25\,]$~GeV). All
matrix elements are first measurements for $-t > 0.6$~GeV$^2$. Moreover, differential cross sections
d$\sigma$/d(cos\,$\Theta_{\rm \,c.m.}^{\,\omega}$) and the corresponding angle-dependent unpolarized spin-density
matrix elements in the Adair frame are presented for the incident photon energy range 1.56--3.80~GeV (corresponding
to $W \in [\,1.95,2.83\,]$~GeV). These new $\omega$~photoproduction data are consistent with earlier CLAS results
but extend the energy range well beyond the nucleon resonance region into the Regge regime. The comparison with
Regge-theory-based model predictions shows that the new data impose more stringent constraints on our understanding
of $\omega$~photoproduction.
\end{abstract}




\newpageafter{author}

\end{frontmatter}




\begin{multicols}{2} 
  
\section{Introduction\label{Introduction}}
The light-flavor vector mesons $\rho$, $\omega$, and $\phi$ have the same spin, parity and charge conjugation
quantum numbers, $J^{PC} = 1^{--}$, as the photon and, therefore, these mesons play an important role
in photoproduction. In high-energy $\omega$~photoproduction, the non-resonant amplitude is dominated by
a diffractive mechanism, whereas single-$\pi$ exchange has been predicted to dominate in the nucleon resonance
regime at lower energies. The contribution from nucleon excitations in $\omega$~photoproduction was recently
studied by several groups~\cite{Akbar:2017uuk,Roy:2017qwv,Roy:2018tjs,Wei:2019imo}. Surprisingly, an amplitude 
analysis within the Bonn-Gatchina framework~\cite{Wilson:2015uoa} found Pomeron-exchange to provide the largest 
contribution to the non-resonant background. In vector-meson photoproduction, the spin observables of the vector 
mesons are represented by their density matrix. Its elements, the spin-density matrix elements (SDMEs), are the 
physical observables that provide information on the underlying production mechanism and the decay distribution, 
$W$(cos\,$\theta, \phi$), of the vector meson. In Ref.~\cite{Wilson:2015uoa}, the $t$-channel amplitude was 
identified by polarized SDMEs using linear beam polarization, and additional beam- and beam-target polarization 
observables~\cite{Klein:2008aa,Eberhardt:2015lwa}.

At low momentum transfer, the differential cross section exhibits the typical exponential $-t$~dependence 
expected from Pomeron and Reggeon exchange. This diffractive production mechanism is described by the incident 
photon that fluctuates into a vector meson and then elastically scatters from the target nucleon. At large 
momentum transfer, the cross section shows a flat behavior that can be explained in terms of QCD-inspired 
two-gluon exchange~\cite{Collins:1983fg}. The impact parameter becomes small enough and prevents the two
constituent gluons from forming the exchanged Pomeron. Moreover, the light, non-strange quark composition
of the $\omega$~meson allows valence quarks to be exchanged between the meson and recoil
nucleon~\cite{Cano:2001sb}. In contrast, the latter mechanism is suppressed in single-$\phi$ meson
photoproduction by the OZI rule and two-gluon exchange dominates. At medium energies, it is important to 
establish factorization of Regge vertices to separate mesons from baryons. This has been discussed in a 
recent Regge-based study that also provides the implication of Regge-pole models on the
SDMEs~\cite{Mathieu:2018xyc}.

The low-energy regime close to threshold, where nucleon resonances dominate $\omega$ photoproduction, and 
the Regge regime are analytically connected. However, the scarcity of $-t$\,-\,dependent cross sections and 
other spin observables has hindered our understanding of the transition from the baryon resonance regime to 
high-energy photoproduction. Moreover, each Reggeon exchange has a known energy behavior, whereas the 
dependence on the momentum exchange is initially unknown and only poorly understood in the transition and 
medium-energy regimes.

In this Letter, we report differential cross sections and various extracted SDMEs for the center-of-mass 
energy range $W\in [\,1.95,3.25\,]$~GeV for the reaction $\vec{\gamma}\,p \to p\,\omega$ using circularly 
polarized tagged photons. The $\omega\to\pi^+\pi^-\pi^0$ decay angular distribution $W$ is parameterized 
by the density matrices $\rho^\alpha,~\alpha = 0,1,2,3$. For circularly polarized incident photons with 
helicities $\lambda_\gamma = \pm 1$, $W^\pm\,({\rm cos}\,\theta,\phi)$ is given by~\cite{Schilling:1969um}
\begin{align}
W^\pm\,&({\rm cos}\,\theta,\phi)  \,=\,  \nonumber\\
 & W^0\,({\rm cos}\,\theta,\phi) \pm \delta_{\,\odot}\,W^3\,({\rm cos}\,\theta,\phi)\,,
\label{Equation:DecayDistribution}
\end{align}
where $\delta_{\,\odot}$ denotes the degree of circular beam polarization, and $\theta$ and $\phi$ are the 
polar and azimuthal angles of the normal to the decay plane ($\hat{k}_{\pi^+}\times \hat{k}_{\pi^-}$) in the 
$\omega$ rest frame. In this analysis, the SDMEs were extracted from the data for each kinematic bin, 
$(W,\,{\rm cos}\,\Theta_{\rm \,c.m.}^{\,\omega})$ or $(W,\,-t)$, by performing an Extended Maximum Likelihood 
fit to Eq.~(\ref{Equation:DecayDistribution}) with
\begin{align}
  \label{Equation:Schilling}
  W&(\theta,\phi,\rho^0)\,=\,\nonumber\\
  & \frac{3}{4\pi}\,\biggl(\frac{1}{2}\,\bigl(1-\rho^0_{00}\bigr)\,
      +\,\frac{1}{2}\,\bigl(3\rho^0_{00}-1\bigr)\,{\rm cos}^2\,\theta \nonumber\\
  &- \sqrt{2}\,{\rm Re}\,\rho^0_{10}\,{\rm sin}\,2\theta\,{\rm cos}\,\phi \nonumber\\
      &-\, \rho^0_{1-1}\,{\rm sin}^2\,\theta\,{\rm cos}\,2\phi\biggr)\\[1ex]
  W(\theta&,\phi,\rho^3)\,=\, \frac{3}{4\pi}\,\biggl(\sqrt{2}\,{\rm Im}\,\rho^3_{10}\,{\rm sin}\,
      2\theta\,{\rm sin}\,\phi \nonumber\\ &+ {\rm Im}\,\rho^3_{1-1}\,{\rm sin}^2\,\theta\,{\rm sin}\,
      2\phi\biggr)\,,
 \end{align}
where $\rho^0_{00}$, $\rho^0_{1-1}$, Re\,$\rho^0_{10}$ and Im\,$\rho^3_{10}$, Im\,$\rho^3_{1-1}$ are the 
unpolarized and polarized SDMEs, respectively.

\section{Previous Measurements\label{PreviousMeasurements}}
The differential cross sections d$\sigma$/d$t$ were initially measured in the 1970s and 1980s at the Stanford 
Linear Accelerator Center (SLAC) at $E_\gamma = 2.8, 4.7, 9.3$~GeV using a laser-produced polarized photon 
beam~\cite{Ballam:1970ex,Ballam:1972eq} and at the Cornell Laboratory of Nuclear Studies using incident 
tagged-photons with energies between 7.5~GeV and 10.5~GeV~\cite{Abramson:1976ks}. At the electron synchrotron 
NINA, Daresbury Laboratory, measurements were made in the photon energy range 2.8 to 4.8~GeV at backward 
angles~\cite{Clifft:1977yi} and by the LAMP2 Group at forward angles~\cite{Barber:1985fr}. The LAMP2 group 
published very wide energy bins. Therefore, these data and earlier statistically limited data from the 
Cambridge~\cite{Cambridge:1966} and DESY~\cite{ABBHHM:1968} bubble chamber groups are omitted here. More 
recently, the CLAS Collaboration at Jefferson Laboratory reported cross section data for four different 
energy bins between 3.2 and 3.92~GeV~\cite{Battaglieri:2002pr}. Unpolarized SDMEs were extracted at 
SLAC~\cite{Ballam:1972eq}, Cornell~\cite{Abramson:1976ks}, and Daresbury~\cite{Barber:1985fr} at the 
above energies for $-t < 0.6$~GeV$^2$, by CLAS~\cite{Williams:2009ab} for $E_\gamma < 3.83$~GeV in a 
$(E_\gamma,\,{\rm cos}\,\Theta_{\rm \,c.m.}^{\,\omega})$ representation, and at ELSA~\cite{Wilson:2015uoa}
below 2.5~GeV. Our measurements are first measurements for $-t > 0.6$~GeV$^2$. Some experiments reported 
data on SDMEs using linear polarization~\cite{Ballam:1972eq,Wilson:2015uoa}. The polarized SDMEs presented 
here have been extracted from photoproduction data using circular polarization; they are also first measurements.

\section{The CLAS-g12 Experiment\label{ExperimentalSetup}}
These new results were obtained within the framework of the CLAS-g12 experiment~\cite{Hu:2020ecf}, conducted 
at Jefferson Laboratory before the 12 GeV upgrade. CLAS-g12 used longitudinally polarized electrons with 
energies of $E_{\rm e^-} = 5.715$~GeV and polarization $\sim$67.2\%~\cite{Akbar:Thesis} on a liquid-hydrogen 
target. Circularly polarized photons were obtained by transferring the polarization from the electrons to the 
photons in a brems\-strahlung process when the electrons scattered off an amorphous gold radiator. The degree 
of the photon-beam polarization, $\delta_{\,\odot}$, was given as a function of the degree of electron-beam 
polarization, $\delta_{\,e^-}$, and the photon-beam energy $E_{\,\gamma}$ as~\cite{Olsen:1959zz}:
\begin{equation}
\delta_{\,\odot}\,=\,\delta_{e^-}\,.\,\frac{4x\,-\,x^2}{4\,-\,4x\,+\,3x^2}~,
\end{equation} 
where $x = E_{\gamma}/E_{\, e^-}$. 

The polarized photons were energy and time tagged with resolutions of 0.1\% and 100~ps, respectively, using 
a photon tagging system~\cite{Sober:2000we}.

The CLAS detector, with its six-fold symmetry about the beamline (sectors), was capable of detecting charged 
particles with a laboratory polar-angle coverage of $[8,\,142]^{\circ}$ and almost $2\pi$ coverage in the 
azimuthal angle~\cite{Mecking:2003zu}. The final-state particles traversed several layers of sub-detectors,
including drift chambers (DC) and time-of-flight (TOF) scintillators. A start counter (SC) provided the initial
time information of the events. For an event to be recorded, the trigger required a scattered electron in the 
bremsstrahlung tagger in coincidence with either (a) (at least) three charged tracks in different sectors with 
no restrictions on photon energy, or (b) only two tracks in different sectors with the additional requirement 
of at least one tagger photon with an energy above 3.6~GeV.

\subsection{Event Selection\label{Event Selection}}
In this analysis, the~$\omega$ was reconstructed from its $\pi^+\pi^-\pi^0$~decay with a branching ratio of 
$(89.3\pm 0.6)\%$~\cite{ParticleDataGroup:2024cfk}. The CLAS spectrometer was designed primarily for the
detection and measurement of charged particles. Nevertheless, the overconstrained event kinematics enabled
the reconstruction of a single undetected neutral meson through the missing-mass technique. Candidate
$\gamma p \to p\pi^+\pi^-(\pi^0)$ events were initially selected by requiring exactly one reconstructed
proton track and two reconstructed charged-pion tracks, with the $\pi^0$ identified as the missing particle.
Positively and negatively charged pions were identified based on the opposite curvatures of their tracks in the
CLAS toroidal magnetic field. The acceptance for $\pi^-$ mesons was lower than that for $\pi^+$ mesons
because negatively charged particles were bent toward the beamline, causing a significant fraction of them
to escape through the forward hole of the CLAS spectrometer. The $\pi^0$ meson was subsequently identified
through kinematic fitting. The selection of the $p\,\pi^+\pi^-\pi^0$ sample is outlined in detail in
Ref.~\cite{Hu:2020ecf}.

In a brief summary, events were selected to have exactly one incident-photon candidate with a timing
(using the photon tagger) at the event vertex within 1~ns of the event time provided by the SC. Only those
events that had exactly one proton, $\pi^+$, and $\pi^-$ were retained. To further improve the particle
identification, each particle's $\beta$~value was calculated separately from its measured momentum using
the DC, $\beta_{\rm \,DC}$, and from its measured velocity using the TOF system and the SC, 
$\beta_{\rm \,TOF}$. Events were selected based on good agreement of $\beta_{\rm \,DC}$ and 
$\beta_{\rm \,TOF}$~\cite{Hu:2020ecf}. The momenta of the final-state particles were corrected 
for energy losses using standard CLAS techniques.

\begin{figure}[H]
  \begin{center}
    \includegraphics[width=0.50\textwidth]{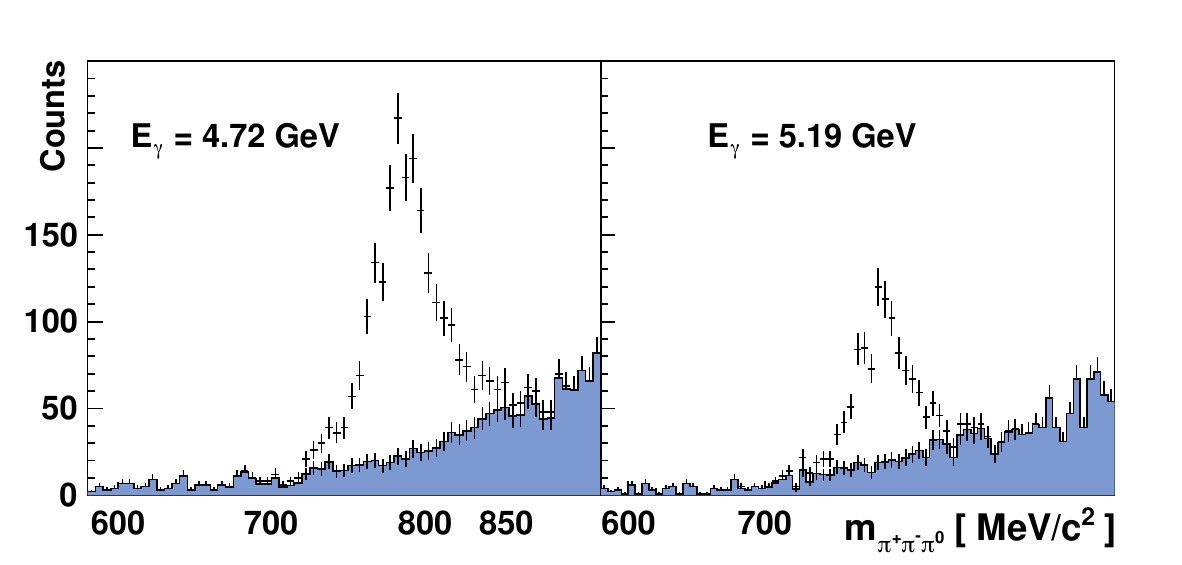}
    \caption{\label{Figure:Masses} Examples of signal and background mass distributions from data after applying 
       all kinematic cuts. The invariant $\pi^+\pi^-\pi^0$ masses are shown for approximately 70-MeV-wide 
       $E_\gamma$~bins with center values of $E_\gamma = 4.72$~GeV (left) and 5.19~GeV (right). See text for details.} 
  \end{center}
\end{figure}

A four-constraint (4C) kinematic fit to the exclusive $\gamma p \to p\,\pi^+\pi^-$ reaction imposing energy 
and momentum conservation aided in tuning the full covariance matrix. The reaction 
$\gamma p \to p\,\pi^+\pi^-\,({\rm missing}\,\pi^0)$ was then kinematically fit, and events with a confidence 
level below $0.01$ were rejected, removing most of the $\pi^+\pi^-$~background.

\subsection{Background Subtraction}
In a final step, the remaining background beneath the $\omega$ peak was accounted for using a multivariate
side-band subtraction technique~\cite{Williams:2008sh}, which determined the probability $Q$ for an event to
be a signal event (as opposed to background) on the basis of a sample of its nearest kinematic neighbors in a
very small region of the multi-dimensional $\pi^+\pi^-\pi^0$ phase space around the candidate
event~\cite{Hu:2020ecf,Williams:2008sh}. The method assumes that the signal and background distributions do
not vary rapidly in the selected region. The $\pi^+\pi^-\pi^0$~mass distribution of each event and its nearest
kinematic neighbors was fit using a Johnson form for the signal probability function (PDF) and a second-order
Chebychev polynomial for the background PDF. The value of $Q$ was then defined as the ratio of signal,~$s(m)$,
to total amplitude,~$s(m) + b(m)$, at the mass~$m$ of the candidate event:
\begin{equation}
  Q \,=\, \frac{s(m)}{s(m) \,+\, b(m)}\,.
  \label{Equation:Q_factor}
\end{equation}

The $Q$-factor method introduces correlations among events and their nearest neighbors because a given
event can contribute to the nearest-neighbor sets of multiple {\it seed} events. The resulting systematic
uncertainty in the extracted $\omega$ yield arising from these correlations, for a given kinematic bin, is
therefore calculated as
\begin{equation}
\sigma^2_{\rm corr}\,=\,\sum_{i,\,j}\,\sigma_Q^i\,\rho_{ij}\,\sigma_Q^j\,,
\label{Correlation}
\end{equation}
where the sum over $i$ and $j$ extends over all events in the bin, $\sigma_Q^i$ and $\sigma_Q^j$ are the
fit uncertainties associated with events $i$ and $j$, respectively, and $\rho_{ij}$ denotes the correlation coefficient
between the two events. The correlation coefficient is defined as the fraction of nearest-neighbor events shared
by events $i$ and $j$. In this analysis, the total uncertainty on the signal yield in each kinematic bin was obtained
by adding the systematic uncertainty given by Eq.~\ref{Correlation} in quadrature with the statistical
uncertainty~\cite{Williams:2008sh}.

Figure~\ref{Figure:Masses} demonstrates the quality of the background subtraction. Examples of signal and 
background distributions are shown in the invariant $\pi^+\pi^-\pi^0$ mass obtained by weighting each event 
with $Q$ and $1-Q$, respectively. Likewise, in this analysis, event yields for cross section measurements were 
based on sums over $Q$~values, whereas contributions to the decay angular distribution given by 
Eq.~(\ref{Equation:DecayDistribution}) have been weighted event-by-event with the corresponding $Q$~values to 
subtract background.

The $\gamma p\to p\omega\to p\pi^+\pi^-\pi^0$ acceptance of CLAS was evaluated in GEANT3~\cite{Brun:1994aa} 
Monte-Carlo simulations by generating events evenly distributed across the available phase space. The Monte 
Carlo events were then analyzed using the same reconstruction and selection criteria that were applied to the 
data events~\cite{Hu:2020ecf}.

\section{Results\label{Results}}
\begin{figure}[H]
  \begin{center}
    \includegraphics[width=0.48\textwidth]{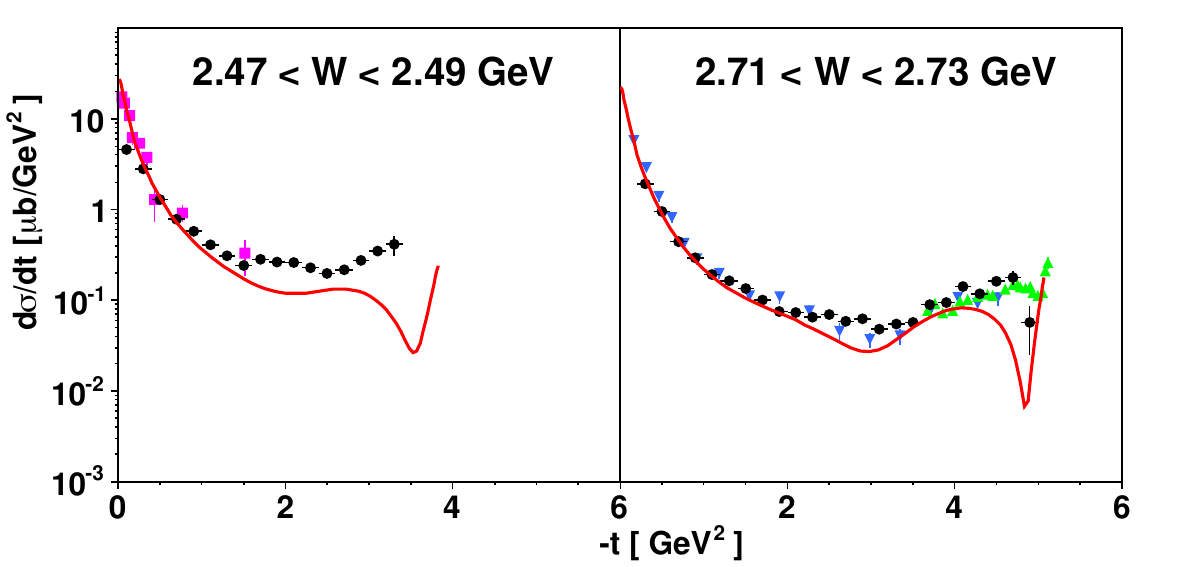}
    \includegraphics[width=0.48\textwidth]{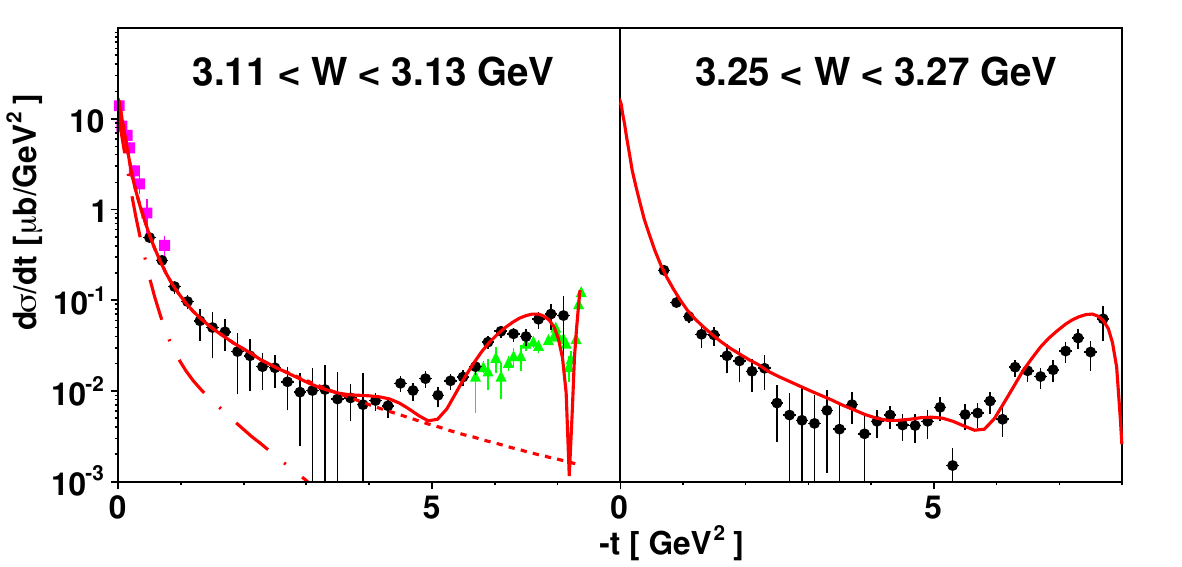}
    \caption{\label{Figure:CrossSections} Four selected examples of d$\sigma$/d$t$ in 20-MeV-wide center-of-mass 
         bins. The new CLAS data are shown as the black solid circles ({\large $\bullet$}) and the uncertainties include 
         the statistical and $Q$-value correlation uncertainties added in quadrature, see text for more details.
         The remaining energy bins for the  entire $W$~range 2.45--3.25~GeV 
         are also available~\cite{SuppMaterial:CrossSections}. Also shown for 
         comparison are earlier data from CLAS~\cite{Battaglieri:2002pr}~({\footnotesize\color{blue}{$\blacktriangledown$}}) 
         for the energy bin $E_\gamma\in [\,3.38,\,3.56]$~GeV, (all published data below 6~GeV from) 
         NINA~\cite{Clifft:1977yi}~({\footnotesize\color{green}{$\blacktriangle$}}) at $E_\gamma = 3.5$~GeV and 4.7~GeV, 
         and SLAC~\cite{Ballam:1972eq}~({\tiny\color{magenta}{$\blacksquare$}}) at $E_\gamma = 2.8$~GeV and 4.7~GeV. In 
         these energy bins, $\Theta_{\rm \,c.m.}^{\,\omega} = 90^\circ~(180^\circ)$ corresponds to 
         $-t = 1.92,\,2.54,\,3.68,\,4.08$~GeV$^2$ ($3.82,\,5.07,\,7.37,\,8.16$~GeV$^2$), respectively. The red curves 
         denote the model predictions of Ref.~\cite{Laget:2021qwq}: full solution (solid curve), only two-gluon and 
         $f_2$~contributions (long dash-dotted curve), additional $\pi$~exchange (dashed line).}
  \end{center}
\end{figure}

The differential $\omega$~photoproduction cross section as a function of $-t$ is shown in 
Fig.~\ref{Figure:CrossSections} for four selected energy bins. The horizontal bars reflect the bin size. 
The full set of cross sections d$\sigma$/d(cos\,$\Theta$) in 20-MeV-wide and 10-MeV-wide bins for the 
$W$~range 1.95--2.35~GeV and 2.35--2.83~GeV, 
respectively, and d$\sigma$/d$t$ in 
20-MeV-wide bins for the $W$~range 2.45--3.25~GeV 
is provided as supplemental 
material~\cite{SuppMaterial:CrossSections}. A ratio distribution of CLAS-g11a~\cite{Williams:2009ab} and 
our new CLAS-g12 results using the d$\sigma$/d(cos\,$\Theta$)~cross sections shows a fairly Gaussian 
distribution, which indicates that an overall decrease of about~1.2\% is observed in these new data 
relative to the previous CLAS data. The earlier CLAS-g6 results~\cite{Battaglieri:2002pr} shown in 
Fig.~\ref{Figure:CrossSections} are also in excellent agreement with these new results. Data from NINA at 
large momentum transfer~\cite{Clifft:1977yi} underestimate the new CLAS results by about 25\%. In the 
region $-t < 1$~GeV$^2$, good agreement is observed with the previous SLAC measurements~\cite{Ballam:1972eq}. 
The d$\sigma$/d$t$ cross sections show the expected exponential $Ae^{Bt}$~behavior for 
$0.1 < -t < 0.5$~GeV$^2$. We have determined $B = 4.6\pm 0.2$~GeV$^{-2}$ and $B = 4.1\pm 0.3$~GeV$^{-2}$
for $W$ = 2.48~GeV and 2.72~GeV, respectively. These values are consistent with the CLAS measurements of
Ref.~\cite{Battaglieri:2002pr}.

\begin{figure*}[t]
  \begin{center}
    \includegraphics[width=0.477\textwidth]{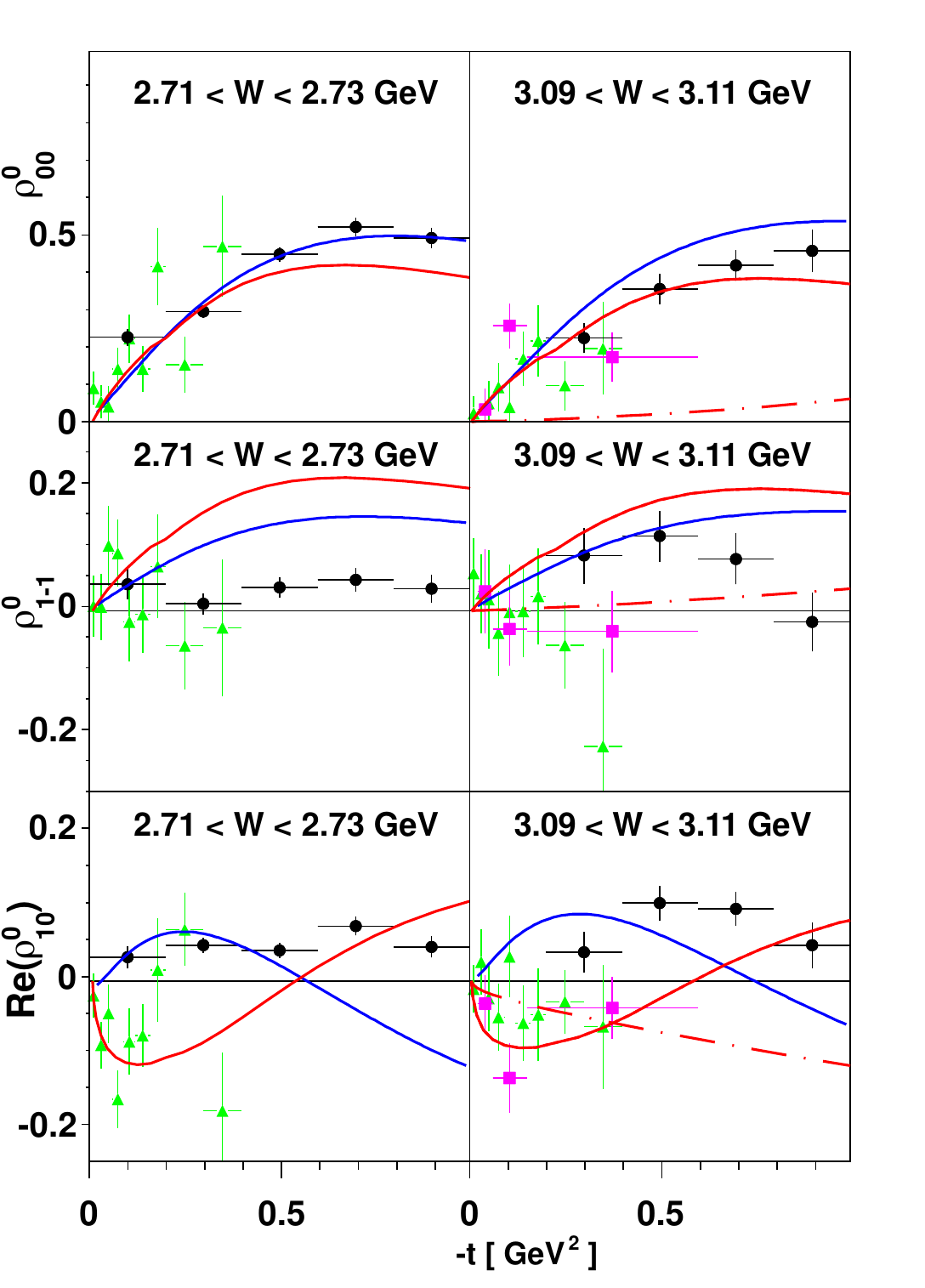}\qquad
    \includegraphics[width=0.46\textwidth]{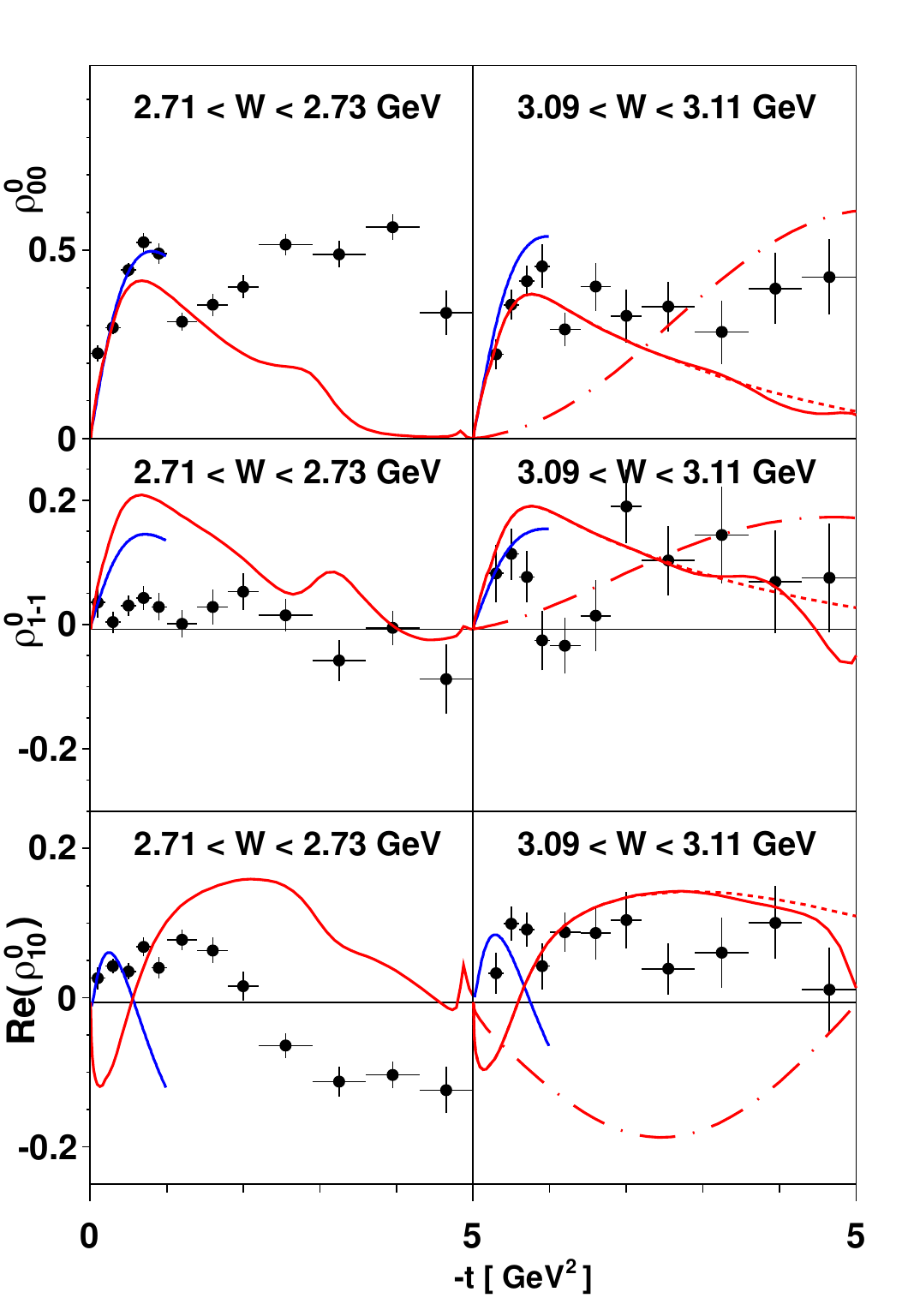}
    \caption{\label{Figure:UnpolarizedSDMEs} Examples of the three unpolarized SDMEs $\rho^0_{00}$ (top),
        Re($\rho^0_{10}$) (middle), $\rho^0_{1-1}$ (bottom) in the Adair frame. The new CLAS data are shown 
        as the black solid circles ({\large $\bullet$}). The remaining energy bins for the entire $W$~range
        2.45--3.25~GeV 
        are also available~\cite{SuppMaterial:SDMEs}. On the left, the three elements 
        are shown for $0 < -t < 1$~GeV$^2$ to allow for better comparison with previously published results from 
        SLAC~\cite{Ballam:1972eq}~({\tiny \color{magenta}{$\blacksquare$}}) at $W \approx 3.11$~GeV and 
        LAMP2~\cite{Barber:1985fr}~({\footnotesize \color{green}{$\blacktriangle$}}) at $W\approx 2.8$~GeV and 
        $W \approx 3.1$~GeV. On the right, the new results are shown for $0 < -t < 5$~GeV$^2$. The blue curves 
        denote the model prediction of Ref.~\cite{Mathieu:2018xyc}; see caption of Fig.~\ref{Figure:CrossSections} 
        for the red curves of Ref.~\cite{Laget:2021qwq}.}
  \end{center}
\end{figure*}

A major contribution to the overall systematic uncertainty comes from the background subtraction (see 
details in Refs.~\cite{Williams:2008sh,Hu:2020ecf}). This absolute contribution is added in quadrature 
to the statistical uncertainty and shown for each data point in Fig.~\ref{Figure:CrossSections}. 
Additional scale-type systematic uncertainties originate from modeling the detector acceptance 
$(7.1\%)$~\cite{Hu:2020ecf,CLAS-NOTE-2017-002}, kinematic fitting $(1.2\%)$, trigger efficiency correction 
$(1.1\%)$, liquid-hydrogen density $(0.5\%)$~\cite{CLAS-NOTE-2017-002}, photon flux normalization 
$(5.7\%)$~\cite{CLAS-NOTE-2017-002}, and the $\omega\to\pi^+\pi^-\pi^0$ branching fraction 
$(0.6\%)$~\cite{ParticleDataGroup:2024cfk}.

\begin{figure}[H]
\includegraphics[width=0.48\textwidth]{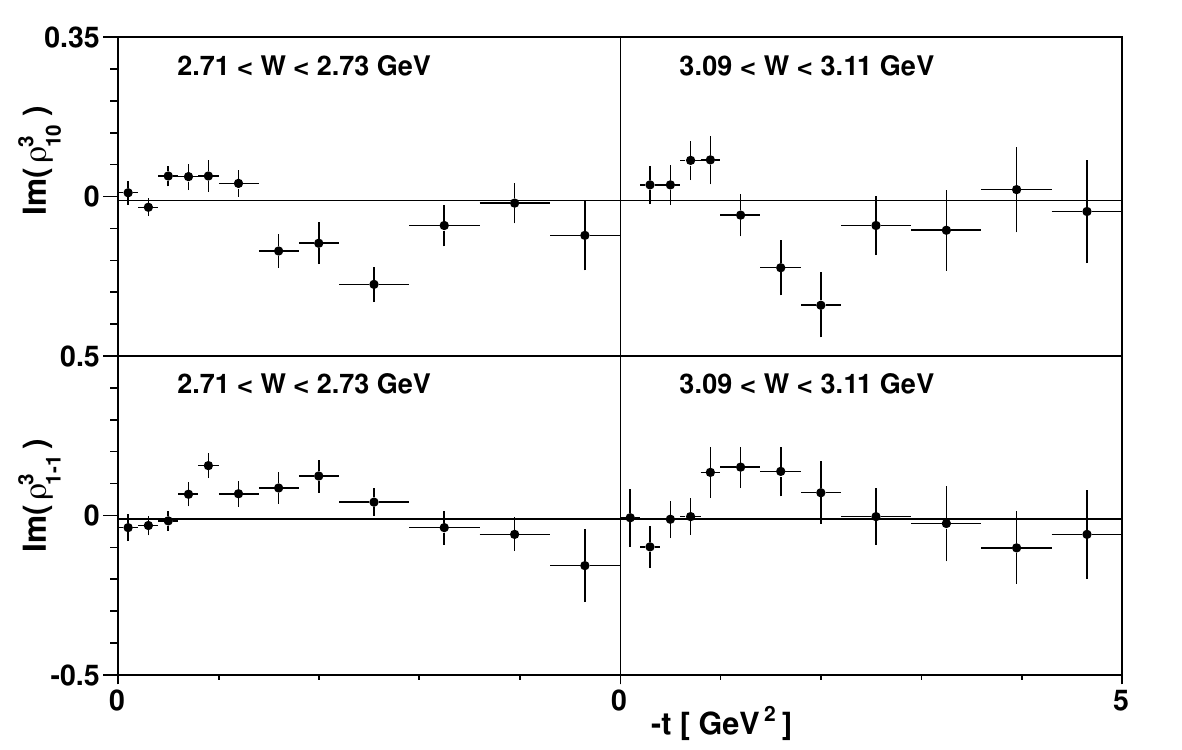}
\caption{\label{Figure:PolarizedSDMEs} Examples of the two polarized SDMEs Im($\rho^3_{10}$) 
  and Im($\rho^3_{1-1}$) in the Adair frame for the same two energy bins presented in
  Fig.~\ref{Figure:UnpolarizedSDMEs}. The remaining energy bins for the entire $W$~range
  2.45--3.25~GeV 
  are also available~\cite{SuppMaterial:SDMEs}. These results are 
  first measurements and therefore, a model description is not yet provided.} 
\end{figure}

For each ($W$,\,cos$\,\Theta^{\,\omega}_{\rm \,c.m.}$) and ($W$,\,$-t$) bin, an event-based 
maximum-likelihood technique was applied to fit Eq.~(\ref{Equation:DecayDistribution}) to the two-dimensional 
$\omega$~decay angular distribution spanned by the two angles of the normal to the three-pion plane in the 
$\omega$~rest frame. The analysis was performed in both the helicity and the Adair frames. Examples of the 
three unpolarized SDMEs are shown in Fig.~\ref{Figure:UnpolarizedSDMEs}~(left) for two selected 20-MeV-wide 
$W$~bins and $-t < 1$~GeV$^2$ to facilitate comparison with previous results. The full $-t$~range for 
these two $W$~bins is shown in Fig.~\ref{Figure:UnpolarizedSDMEs}~(right) and for the two  polarized
SDMEs in Fig.~\ref{Figure:PolarizedSDMEs}. The full set of SDMEs from this analysis is available in
Ref.~\cite{SuppMaterial:SDMEs}. The observables are presented in the Adair frame to facilitate the
comparison with earlier CLAS data~\cite{SuppMaterial:SDMEs}. The main advantage of this frame is that
the $z$-axis is defined with respect to the beam direction in the overall center-of-mass (CM) frame, making
the Adair frame particularly natural for studying the reaction dynamics. Systematic uncertainties were determined
for all SDMEs by modifying each $Q$~factor by its corresponding fit uncertainty $\sigma_Q$ and re-extracting
the observables. The absolute difference was taken as the systematic uncertainty and added in quadrature to
the statistical uncertainty.

The Regge-based model of Ref.~\cite{Mathieu:2018xyc} was developed to understand SDMEs in light 
vector-meson photoproduction and to study the $\gamma$\,-\,$\omega$ vertex. In the model, the Reggeon
amplitude factorizes into a product of two vertices, which describe the photon and proton interactions. 
This factorization follows from unitarity in the $t$~channel and allows for the study of the helicity 
structure at the photon vertex independently from the target. In the center-of-mass frame, the net 
helicity transfer between the $\omega$~meson and the photon $|\lambda_\gamma - \lambda_\omega|$ can be 
0, 1, or 2, and is referred to as helicity conserving, single- and double-helicity flip, respectively. 
Both helicity-flip couplings have been fit to SLAC data~\cite{Ballam:1972eq} at an energy 
of $E_\gamma = 9.3$~GeV and extrapolated to lower energies. The model is shown as the blue curve in
Fig.~\ref{Figure:UnpolarizedSDMEs}. In addition to natural-parity $a_2$ and $f_2$~exchanges, as well as
unnatural-parity $\pi$~exchange (naturality given as $\eta = P(-1)^J$ with parity $P$), natural-parity
Pomeron exchange has been considered. The contribution of $\eta$~exchange is found to be negligible
in $\omega$~photoproduction and axial-vector trajectories have been neglected in the
model~\cite{Mathieu:2018xyc} since the pseudoscalar exchanges are sufficient to describe the unnatural
components of the SDMEs. The model compares well with the unpolarized SDME $\rho^0_{00}$ within its
valid range of about $0 < -t < 1$~GeV$^2$, but overestimates the SDME $\rho^0_{1-1}$ and even exhibits
the wrong sign for ${\rm Re}\,\rho^0_{10}$ due to an overestimated contribution of unnatural $\pi$~exchange
relative to natural Pomeron and tensor $a_2/f_2$~exchanges. In the model, a reduction of $\pi$~exchange
would flip the sign of $\rho^0_{10}$ to match data. 

The model described in Ref.~\cite{Laget:2019tou} was developed to analyze the cross sections of meson 
photo- and electroproduction channels. The red curves in Figs.~\ref{Figure:CrossSections}
and~\ref{Figure:UnpolarizedSDMEs} show the most recent predictions~\cite{Laget:2021qwq}. In the
$\phi$~channel, the two-gluon exchange contribution reproduces the production cross
section~\cite{CLAS:2000kid} and the SDME $\rho^0_{00}$ at high energy fairly well~\cite{CLAS:2003pux}.
In $\omega$~production, this contribution is supplemented by $\pi$ and $f_2$ exchanges at forward angles, 
as well as proton exchange at backward angles. While the relativistic spin-momentum dependence of the 
$\pi$~exchange amplitude is fully taken into account, the corresponding dependence of the $f_2$~exchange 
amplitude is chosen to be the same as that for two-gluon exchange ($f_2$-Pomeron equivalence) and, 
consequently, both conserve helicity. The only difference resides in the propagator: The two-gluon 
amplitude is purely imaginary (absorptive), whereas the $f_2$ amplitude is real and incorporates the 
Regge propagator with the linear trajectory of the $f_2$. At intermediate angles, the use of a saturating
Regge trajectory for the $\pi$ meson is necessary to account for previously measured cross
sections~\cite{Cano:2001sb}. As can be seen in Fig.~\ref{Figure:CrossSections}, the model agreement with 
the cross sections is excellent (without any adjustments), especially at the highest energies, where 
the data seem to indicate the predicted minimum at the most backward angles. In contrast, the new SDMEs 
reported here are not well described. To improve the model, it will be necessary to go beyond the 
Pomeron-$f_2$ equivalence, and take into account the actual spin-momentum structure of the $f_2$~exchange 
amplitude. This approach is not expected to change the description of the differential cross sections. 
Furthermore, a saturating trajectory for the $f_2$ can be used, at the expense of modifying that for 
$\pi$~exchange.

\section{Summary and Conclusions\label{Summary}}
In summary, the first comprehensive data set of photoproduction cross sections and (un)polarized SDMEs 
has been presented for the reaction $\gamma p\to p\omega$ using circularly polarized tagged photons up 
to $W = 3.25$~GeV and the CLAS spectrometer at Jefferson Laboratory. The results are given in terms of
cos\,$\Theta_{\rm \,c.m.}^{\,\omega}$ for $W\in [\,1.95,\,2.83\,]$~GeV and in terms of $-t$ for $W\in 
[\,2.45,\,3.25\,]$~GeV. We observe excellent agreement with predictions for the cross sections. At the 
highest energy, the data seem to be consistent with a downward bending at large $|-t|$~values providing 
a hint for the node predicted by the model in Ref.~\cite{Laget:2021qwq}. Moreover, the new set of SDMEs 
shows fair agreement with the predictions of the JPAC model~\cite{Mathieu:2018xyc} for $-t < 0.5$~GeV$^2$. 
However, the data indicate that the model significantly overestimates the contribution of unnatural to 
natural $\pi$~exchange, which is most obvious for ${\rm Re}\,\rho^0_{10}$. These new measurements
provide a testing ground for the improvements required in the current models. In particular, the first extraction
of the polarized SDMEs Im\,$\rho^3_{10}$ and Im\,$\rho^3_{1-1}$ provides important additional information to 
improve our understanding of $\omega$ photoproduction in the transition from the baryon resonance regime 
to high-energy photoproduction.

\section*{Acknowledgments\label{Acknowledgments}}
The authors gratefully acknowledge the excellent support of the technical staff at Jefferson 
Lab and all participating institutions. This research is based on work supported by the U.\,S. 
Department of Energy, Office of Science, Office of Nuclear Physics, under Contract No. 
DE-AC05-06OR23177. The group at Florida State University acknowledges additional support from 
the U.S. Department of Energy, Office of Science, Office of Nuclear Physics, under Contract No. 
DE-FG02-92ER40735. This work was also supported by the U.\,S. National Science Foundation, the 
State Committee of Science of Republic of Armenia, the Italian Istituto Nazionale di Fisica Nucleare, 
the French Centre National de la Recherche Scientifique, the French Commissariat a l’Energie Atomique, 
the Scottish Universities Physics Alliance (SUPA), the United Kingdom Science and Technology Facilities
Council (STFC), the National Research Foundation of Korea, the Deutsche Forschungsgemeinschaft
(SFB/TR110), the Russian Science Foundation under Grant No. 16-12-10267, the National Natural 
Science Foundation of China under Grants No. 11475181 and No. 11635009, and in part by the Chilean 
National Agency of Research and Development ANID PIA/APOYO AFB180002. V.~Mathieu is a Serra
H\'unter fellow and acknowledges support from the Spanish national Grant No. PID2019–106080 GB-C21
and PID2020-118758GB-I00.


\end{multicols} 

\end{document}